\documentclass[prb,reprint]{revtex4-1} 
\usepackage{amsmath}
\usepackage{amssymb}
\usepackage{graphicx}
\usepackage{enumitem}

\newcommand{\dt}{\Delta t}
\newcommand{\dv}{\Delta v}

\usepackage{tikz}
\usetikzlibrary{calc,trees,positioning,arrows,chains,shapes.geometric,  decorations.pathreplacing,decorations.pathmorphing,shapes,  matrix,shapes.symbols}

\begin{document}
\title{Computational modeling of diffusive dynamics in a bouncer system 
with an irregular surface}
\author{Luiz Antonio Barreiro}
\email{luiz.a.barreiro@unesp.br}

\address{Universidade Estadual Paulista\\
Instituto de Geoci\^encias e Ci\^encias Exatas\\
Departamento de Física}

\begin{abstract}
The horizontal dynamics of a bouncing ball  interacting with an irregular surface is investigated and is found to demonstrate behavior analogous to a random walk. Its stochastic character is substantiated by the calculation of a permutation entropy. The probability density function associated with the particle positions evolve to a Gaussian distribution, and the second moment follows a power-law dependence on time indicative of diffusive behavior. The results  emphasize that   deterministic systems with complex geometries or nonlinearities can generate behavior that is statistically indistinguishable from random. Several problems are suggested to extend the analysis.
\end{abstract}
\maketitle

\section{Introduction}

Diffusion is a fundamental physical process that  describes the movement of particles from regions of higher
concentration to regions of lower concentration due to random motion.
This process is not only a key concept in physics but is also crucial
in chemistry, biology, and materials science. From the spreading of
ink in water to the exchange of gases in the lungs, diffusion is an
essential mechanism that governs numerous natural and technological
phenomena.\cite{Crank,berg1993random}

The study of diffusion offers an excellent
opportunity to introduce students to core concepts in statistical
mechanics and transport phenomena. The mathematical framework
of diffusion is often described by Fick's laws, first formulated in
the 19th century, which provide a quantitative understanding of how
particles spread over time.\citep{fick1855} Furthermore, the study
of Brownian motion, first observed by Robert Brown in 1827 and later
explained by Albert Einstein in 1905, presents an engaging way to
connect diffusion with the kinetic theory of matter and stochastic
processes.\citep{einstein1905}

Beyond its applications in physics, diffusion is a critical topic
across multiple disciplines. In chemistry, it governs reaction kinetics
and mass transport in solutions and gases.\citep{atkins2018physical}
In biological systems, diffusion enables essential processes such
as oxygen transport and cellular respiration.\citep{alberts2014molecular}
In engineering and materials science, diffusion influences the behavior
of semiconductors, corrosion rates, and the efficiency of batteries
and fuel cells.\citep{shewmon2016diffusion} Additionally, diffusion
plays a crucial role in atmospheric science, helping to explain pollutant
dispersion and climate dynamics.\citep{seinfeld2016atmospheric}

Due to its interdisciplinary significance, diffusion serves as a valuable
framework for physics education, offering a natural bridge between
theoretical concepts and real-world phenomena. This work examines
the foundational principles of diffusion, with an emphasis on its
mathematical formulation and computational modeling as effective pedagogical
tools for deepening students\textquoteright{} understanding of transport
processes.

To explore diffusive behavior without resorting to stochastic techniques
such as random number generation, we investigate a deterministic system
based on classical free fall. Although the dynamics of a falling particle under
gravity are well understood, additional complexity arises when the
particle interacts with a structured surface. The coupling between the 
vertical descent and lateral displacement introduces rich dynamical
behavior that can lead to statistical dispersion.

We focus on a particularly illustrative scenario in which a particle
undergoes successive collisions with a sinusoidal surface. This setup
induces a progressive spreading of the particle's horizontal position
over time. In the following we will simulate this motion, analyze
the resulting distribution of the horizontal displacements, and
highlight key features associated with chaotic dynamics and 
diffusion, thus providing insights into the transition from simple mechanical
motion to complex transport phenomena.

\section{The model}

\begin{figure}[h]
\begin{centering}
\includegraphics[width=0.7\columnwidth]{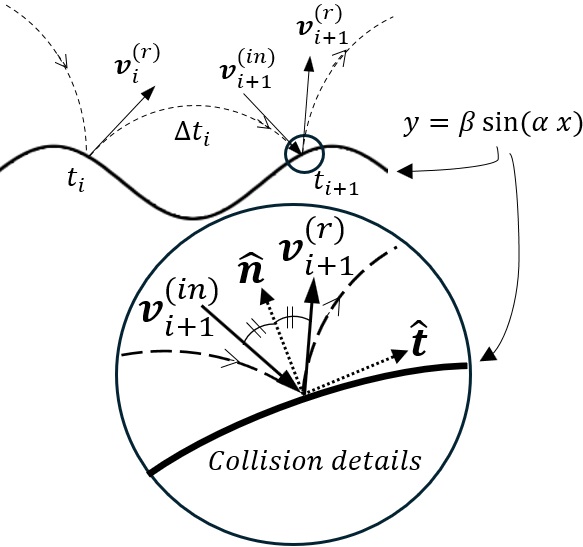}
\par\end{centering}
\caption{The collision details are shown in the  circular region, including
the normal and tangential vectors to the curve at  impact. The dashed
line illustrates the trajectory of the particle, and the collision
 and their corresponding normal vectors are indicated by dotted
arrows. The law of reflection relates the incident velocity
vector $\boldsymbol{v}_{i}^{(\rm in)}$ to the reflected velocity vector
$\boldsymbol{v}_{i}^{(r)}$ as in Eq.~\eqref{map1}.}
\label{FigCollision-1}
\end{figure}

We follow the approach presented in Ref.~\onlinecite{Boscolo2023} and
consider a particle falling on a non-flat surface 
described by 
\begin{equation}
y=\beta\left(\sin(\alpha x)+1\right).\label{curve}
\end{equation}
The particle undergoes successive collisions with the surface, being
reflected at each impact and subsequently following a parabolic
trajectory until the next encounter. In the absence of dissipation,
this process repeats indefinitely. To analyze this sequence of collisions,
it is essential to determine the spatial coordinates of the impact
points as well as the corresponding velocity vectors at the moments
of contact. The reflection velocity vector at each collision is obtained
by considering the local slope of the surface at the point of
impact. The motion can be analyzed  graphically as depicted in Fig.~\ref{FigCollision-1}.

Following the $i$th collision at   $(x_{i},y_{i})$, the
particle is reflected with velocity $\boldsymbol{v}_{i}^{(r)}=\left(v_{x_{i}}^{(r)},v_{y_{i}}^{(r)}\right)$
and subsequently follows a ballistic trajectory under the influence
of gravity until it reaches the next   impact $(x_{i+1},y_{i+1})$.
The corresponding time of flight for this trajectory is denoted by
$\dt_{i,i+1}=t_{i+1}-t_{i}$, such that the total elapsed time
after $N_{\rm col}$  collisions is given by $t_{N_{\rm col}}=\sum_{i=1}^{N_{\rm col}}\dt_{i-1,i}$,
with the initial condition $t_{0}=0$. 

The position of the particle during the free-flight phase evolves
according to
\begin{align}
x_{i+1} & =  x_{i}+v_{x_{i}}^{(r)}\,\dt_{i,i+1},\label{eq:x}\\
y_{i+1} & =  y_{i}+v_{y_{i}}^{(r)}\,\dt_{i,i+1}-\frac{g}{2}\left(\dt_{i,i+1}\right)^{2}.\label{eq:y}
\end{align}
These expressions are evaluated iteratively. To advance the simulation,
it is necessary to determine the time of flight $\dt_{i,i+1}$,
which is obtained by imposing the condition that the coordinate $y_{i+1}$
of the subsequent collision  lies on the surface profile defined
by Eq.~(\ref{curve}). Under this constraint, Eq.~\eqref{eq:y} for $y_{i+1}$ leads to the  transcendental equation 
\begin{eqnarray}
\beta\sin\left[\alpha\left(x_{i}+v_{x_{i}}^{(r)}\,\dt_{i,i+1}\right)\right]&=&
\beta\sin(\alpha x_{i})+v_{y_{i}}^{(r)}\,\dt_{i,i+1} \nonumber \\
&&-\frac{g}{2}\left(\dt_{i,i+1}\right)^{2},\label{eq:Transc}
\end{eqnarray}
which must be solved numerically to determine $\dt_{i,i+1}$,
given the known position $x_{i}$ and the reflected velocity vector
at the $i$th collision. The Appendix outlines
the secant method for solving
this transcendental equations numerically.

In the limit $\beta\ll1$, we can   assume that
the height of the particle at a collision  satisfies $y_{i}\simeq y_{i+1}\simeq0$, while the local slope of the surface may still be nonzero. This
approximation eliminates the need to solve Eq.~\eqref{eq:Transc},
significantly simplifying the analysis. For $\beta\rightarrow0$,
Eq.~\eqref{eq:Transc} reduces to an algebraic form, and the flight
time is readily obtained as
\begin{equation}
\dt_{i,i+1}=\frac{2v_{y_{i}}^{(r)}}{g}.\label{eq:timeSpend}
\end{equation}

To determine the post-collision velocity $\boldsymbol{v}_{i}^{(r)}$,
it is assumed that the tangential component relative to the surface
remains unchanged, while the normal component reverses sign.  At the instant
of collision, the law of reflection relating the incident velocity
vector $\boldsymbol{v}_{i}^{(\rm in)}$ to the reflected velocity vector
$\boldsymbol{v}_{i}^{(r)}$ is
\begin{equation}
\boldsymbol{v}_{i}^{(r)}=\left(\boldsymbol{v}_{i}^{(\rm in)}\cdot\hat{\boldsymbol{t}}_{i}\right)\hat{\boldsymbol{t}}_{i}-\gamma\left(\boldsymbol{v}_{i}^{(\rm in)}\cdot\hat{\boldsymbol{n}}_{i}\right)\hat{\boldsymbol{n}}_{i}, \label{map1}
\end{equation}
where $\hat{\boldsymbol{t}}_{i}$ and $\hat{\boldsymbol{n}}_{i}$
are the unit tangent and normal vectors, respectively. Inelastic collisions
can be taken into account if the dissipative  factor $\gamma$ is included
in the normal component of the reflected velocity. For
 elastic collisions,   $\gamma=1$.

From Fig.~\ref{FigCollision-1},
the unit tangent and normal vectors can be expressed as follows 
\begin{equation}
\hat{\boldsymbol{t}}_{i}=\frac{1}{\sqrt{1+\lambda_{i}^{2}}}\left[\begin{array}{c}
1\\
\lambda_{i}
\end{array}\right],\quad\hat{\boldsymbol{n}}_{i}=\frac{1}{\sqrt{1+\lambda_{i}^{2}}}\left[\begin{array}{c}
-\lambda_{i}\\
1
\end{array}\right],
\end{equation}
where $\lambda_{i}$ is the local slope of the surface, which
from  Eq.~(\ref{curve}) is given by
\begin{equation}
\lambda_{i}=\left.\frac{dy}{dx}\right\lfloor _{x_{i}}=\alpha\beta\cos\left(\alpha x_{i}\right).\label{slope}
\end{equation}
 The velocity vector incident at collision $i+1$, is related to the velocity vector reflected at the previous collision 
$i$ as
\begin{equation}
\boldsymbol{v}_{i+1}^{(\rm in)}=\left[\begin{array}{c}
v_{x_{i}}^{(r)}\\
v_{y_{i}}^{(r)}-g\,\dt_{i,i+1}
\end{array}\right].
\end{equation}
Therefore, the reflected velocity vector in Eq.~(\ref{map1}) takes
the form
\begin{equation}
\boldsymbol{v}_{i+1}^{(r)}=\left[\begin{array}{c}
\dfrac{1-\lambda_{i+1}^{2}}{1+\lambda_{i+1}^{2}}v_{x_{i}}^{(r)}+\dfrac{2\lambda_{i+1}}{1+\lambda_{i+1}^{2}}\left(v_{y_{i}}^{(r)}-g\,\dt_{i,i+1}\right)\\
\dfrac{2\lambda_{i+1}}{1+\lambda_{i+1}^{2}}v_{x_{i}}^{(r)}-\dfrac{1-\lambda_{i+1}^{2}}{1+\lambda_{i+1}^{2}}\left(v_{y_{i}}^{(r)}-g\,\dt_{i,i+1}\right)
\end{array}\right].\label{map2}
\end{equation}

Note that, regardless of whether the flight time
$\dt_{i,i+1}$ is determined using the transcendental equation,
Eq.~\eqref{eq:Transc}, or the simplified expression,
Eq.~\eqref{eq:timeSpend}, its value depends solely
on the position and velocity at the $i$th collision. For notational convenience, we define $\tau_{i}\equiv\dt_{i,i+1}$,
and index the flight time using a single subscript.

Similarly, the local slope at the subsequent collision, $\lambda_{i+1}$,
can be expressed as a function of quantities determined at the $i$th
collision: 
\begin{equation}
\lambda_{i+1}=\alpha\beta\cos\left[\alpha\left(x_{i}+v_{x_{i}}^{(r)}\,\tau_{i}\right)\right]\equiv\tilde{\lambda}_{i},\label{eq:LambdaTil}
\end{equation}
 where $\tilde{\lambda}_{i}$ denotes the predicted slope at the next
impact, based solely on the information available at step $i$. 

W combine Eqs.~(\ref{eq:x}) and (\ref{map2})  and write
\begin{equation}
\chi_{i+1}=\mathcal{\varPhi}(\chi_{i}), \label{MAP3}
\end{equation}
where 
\begin{equation}
\chi_{i}=\left[\begin{array}{c}
x_{i}\\
v_{x_{i}}^{(r)}\\
v_{y_{i}}^{(r)}
\end{array}\right], \label{MAP3A}
\end{equation}
and
\begin{equation}
\mathcal{\varPhi}(\chi_{i})=\left[\begin{array}{c}
x_{i}+v_{x_{i}}^{(r)}\,\tau_{i}\\
\dfrac{1-\tilde{\lambda}_{i}^{2}}{1+\tilde{\lambda}_{i}^{2}}v_{x_{i}}^{(r)}+\dfrac{2\tilde{\lambda}_{i}}{1+\tilde{\lambda}_{i}^{2}}\left(v_{y_{i}}^{(r)}-g\,\tau_{i}\right)\\
\dfrac{2\tilde{\lambda}_{i}}{1+\tilde{\lambda}_{i}^{2}}v_{x_{i}}^{(r)}-\dfrac{1-\tilde{\lambda}_{i}^{2}}{1+\tilde{\lambda}_{i}^{2}}\left(v_{y_{i}}^{(r)}-g\,\tau_{i}\right)
\end{array}\right]. \label{MAP3B}
\end{equation}

\begin{figure}
\centering\noindent
\tikzstyle{block} = [rectangle, draw, text width=15em, text centered, rounded      corners, minimum height=3em]
\begin{tikzpicture}
 [node distance=2.0cm,
 start chain=going below,]
\node (n1) at (0,0) [block]  {Initial Conditions $\chi_{0}=(x_{0},v_{x_{0}}^{(r)},v_{y_{0}}^{(r)})$};
\node (n2) [block, below of=n1] {Solve  Eq.~(\ref{eq:Transc}) or    Eq.~(\ref{eq:timeSpend}) to obtain $\tau_{i}$};
\node (n3) [block, below of=n2] {Find local inclination $\tilde{\lambda}_{i}$, Eq.~(\ref{eq:LambdaTil})};
\node (n4) [block, below of=n3] {Iterate the map to obtain $\chi_{i+1}$, Eq.~(\ref{MAP3})};

\draw [->] (n1) -- (n2);
\draw [->] (n2) -- (n3);
\draw [->] (n3) -- (n4);

\draw [->] (n4.east) -| ++(1,0) |- (n2.east);

\end{tikzpicture}
\caption{Flowchart for solving the set of equations that determine the sequence
of collisions between the particle and the surface.}
\label{Fig:Flowchart}
\end{figure}
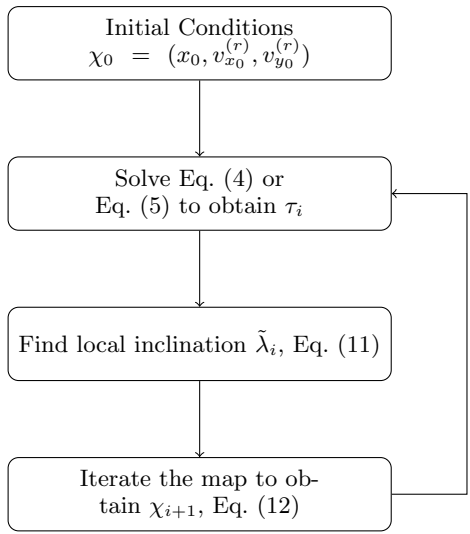

Equations~(\ref{MAP3})--\eqref{MAP3B}, in conjunction with the auxiliary
relations given by Eq.~(\ref{eq:LambdaTil}) and either Eq.~(\ref{eq:Transc})
or Eq.~(\ref{eq:timeSpend}), constitutes a set of equations that must
be solved by an iterative procedure, as outlined in the flowchart in Fig.~\ref{Fig:Flowchart}.

\section{Iterative Process}

\subsection{Initial conditions and Fixed points}

\begin{figure}[h]
\begin{centering}
\includegraphics[width=0.95\columnwidth]{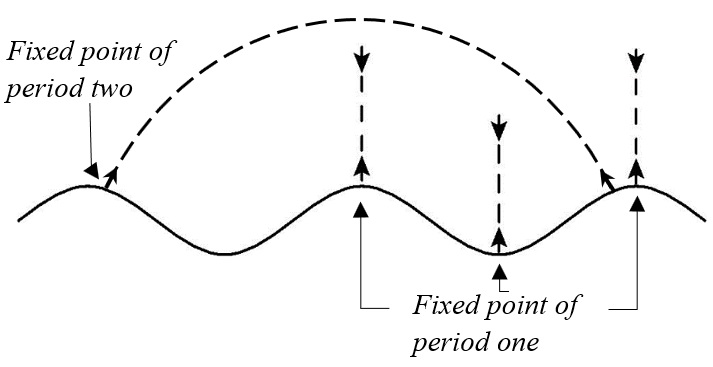}
\par\end{centering}
\caption{\label{fig:two}Fixed points of period one and period two.}
\end{figure}

To study diffusive processes in this dynamical system, we will consider an ensemble
of identical and non-interacting particles by considering different values of the
initial conditions. Because there is no interaction between the particles,
each new initial condition represents a new particle in the system.
We consider  initial conditions such that
all the particles in the ensemble have the same energy. Because the
particles are identical, it is sufficient that the choice of $x_{0}$
and $\boldsymbol{v}_{i}^{(r)}$ keeps the energy per unit mass constant,
\begin{equation}
\frac{E_{0}}{m}=\frac{1}{2}\left[\left(v_{x_{0}}^{(r)}\right)^{2}+\left(v_{y_{0}}^{(r)}\right)^{2}\right]+g\beta\left(\sin(\alpha x_{0})+1\right). \label{Energy}
\end{equation}
Furthermore, it is sufficient to consider initial conditions corresponding
to purely vertical velocities, i.e., $v_{x_{0}}^{(r)}=0$, because
the horizontal displacements necessary for particle diffusion are
induced by the surface irregularities. Lateral diffusion will be suppressed
only in the special case where the initial condition corresponds to
a fixed point of the surface profile, such as a local maximum
or minimum, where the local slope vanishes  and the horizontal
component of the initial velocity is also zero.
These conditions correspond to the period-one fixed points of the dynamical system.
In addition to period-one fixed points, the system also admits higher-order periodic
points, such as period-two fixed points, for which the trajectory
returns to its initial state after two collisions. Representative
examples of such points are illustrated in Fig.~\ref{fig:two}.

To avoid the fixed points,  we choose $x_{0}$ in the range  $-\pi/\alpha$
to $\pi/\alpha$, excluding the extreme values and also the value
 $x_{0}=0$. Thus 
\begin{equation}
x_{0}=\left(-\frac{\pi}{\alpha},\frac{\pi}{\alpha}\right)\setminus\{0\}
\end{equation}
The value of $v_{y_{0}}^{(r)}$ is chosen to keep $E_{0}/m$
constant. We use Eq.~(\ref{Energy}) to obtain 
\begin{equation}
v_{y_{0}}^{(r)}=\sqrt{2\left[\frac{E_{0}}{m}-g\beta\left(\sin(\alpha x_{0})+1\right)\right]}.
\end{equation}

\begin{figure}[h]
\centering
\centering{}\includegraphics[width=0.95\columnwidth]{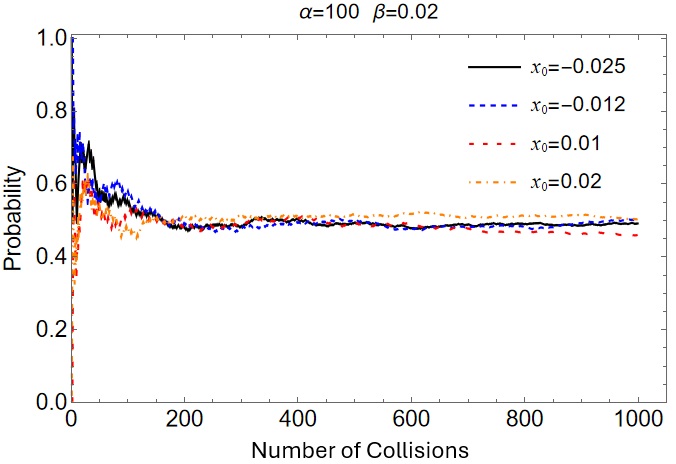}\caption{The evolution of the probability of rightward
motion obtained using Eq.~\eqref{eq:timeSpend}.\label{fig:EvolProb}}
\end{figure}

\subsection{Rightward Jump Probability}

In this preliminary analysis, we examine the distribution of rightward
jumps relative to the total number of jumps. This analysis allows us to estimate
the probability of rightward motion as a function of the number of collisions, which serves as a discrete analogue of time. The simulations
were conducted using the  parameters  
$g=9.8\,\mathrm{m/s}^{2}$, geometric coefficient $\alpha=100\,\mathrm{m}^{-1}$,
surface modulation amplitude $\beta=0.02\,\mathrm{m}$, and energy-to-mass
ratio $E/m=60\,\mathrm{J/m}$.

Given the relation $E/m=gh=60\,\mathrm{J/m}$, the corresponding vertical
height is  $h\approx6.1\,\mathrm{m}$. In contrast, the
 value  $\beta=0.02\,\mathrm{m}$ implies a maximum surface
height of $2\beta=0.04\,\mathrm{m}$, which is about 150 times smaller
than the total energy-equivalent height. This significant disparity
justifies the use of the simplified flight-time equation, Eq.~\eqref{eq:timeSpend}.
If high-precision trajectory tracking is required for individual
particles, the complete transcendental relation must be solved. For
statistical analyses over an ensemble of initial conditions, the simplified
expression provides an adequate approximation.

To investigate the evolution of rightward motion in the system, four independent simulations with different initial
initial conditions were used:   
\begin{equation}
x_{0}\in\{-0.025\,\mathrm{m},\ -0.012\,\mathrm{m},\ 0.010\,\mathrm{m},\ 0.020\,\mathrm{m}\},
\end{equation}
and the simulation was performed for  $N_{\rm col}=1000$ collisions. The number of jumps to the right  after $i$ collisions for each initial condition $j$ 
is  
defined by
\begin{equation}
\mathcal{N}_{+}^{(j)}(i)=\sum_{\ell=1}^{i}\Theta\left(v_{x_{\ell}}^{(j)}\right),
\end{equation}
where $\Theta$ is the Heaviside step function.  Then 
the relative frequency of rightward motion is
\begin{equation}
P_{+}^{(j)}(i)=\frac{\mathcal{N}_{+}^{(j)}(i)}{i}.
\end{equation}
The time series, $\{P_{+}^{(j)}(i)\}_{i=1}^{1000}$, represents the
evolution of the empirical probability of rightward displacement at collision $i$ for
the initial condition $j$.

The results  presented in Fig.~\ref{fig:EvolProb} indicate that the rightward jump probability asymptotically
converges to $0.5$. In other words, as the number of collisions increases,
the system exhibits statistically symmetric behavior: the particle
is equally likely to jump to the left or to the right at each step. This
limiting behavior is characteristic of a one-dimensional symmetric
random walk. Notably, although the system's dynamics are entirely
deterministic, the presence of the undulating surface introduces effective
irregularities that mimic stochasticity.

This emergent statistical behavior, which is manifested in the convergence
of rightward jump counts toward a binomial distribution centered at
$0.5$ suggests that deterministic chaos 
can produce ensemble-level properties that resemble those of genuinely
stochastic processes. Such findings emphasize the capacity of deterministic
systems with complex geometries or nonlinearities to generate outcomes
that are statistically indistinguishable from random behavior. This
result contributes to the broader understanding of how 
chaos can bridge the conceptual gap between deterministic and stochastic
descriptions of dynamical systems.   

\subsection{Stochasticity of the collision force}

Except for special initial conditions (fixed points), the particles
inevitably exhibit diffusion along the $x$-direction. This diffusion
arises from the interaction with the surface, where the collision force
is assumed to remain approximately constant over the short duration
of impact. Due to the irregular structure of the surface, the collision
force $\boldsymbol{F}_{\rm col}$ generally possesses both vertical and
horizontal components. It is straightforward to recognize that the horizontal
component varies in both magnitude and direction at each impact. The
horizontal component of the collision force per unit mass can be expressed
as
\begin{equation}
f_{i}=\frac{v_{x_{i}}^{(r)}-v_{x_{i-1}}^{(r)}}{\tau}=\frac{\dv_{x_{i}}}{\tau},
\end{equation}
where
\begin{equation}
\dv_{x_{i}}=v_{x_{i}}^{(r)}-v_{x_{i-1}}^{(r)}
\end{equation}
 represents the variation in the horizontal velocity between two successive
collisions, and $\tau$ denotes the (very short) duration of the collision.
The quantity $\dv_{x_{i}}$ thus characterizes the evolution
of the horizontal force during the iterative process. Given that
the iterative process yields the sequence of horizontal velocities
$\{v_{x_{0}},v_{x_{1}},\ldots,v_{x_{i},\ldots}\}$,  we can    characterize the system's
dynamics by
the evolution of the horizontal force.

 We can quantitatively determine the degree of stochasticity of the horizontal force by computing the permutation entropy, originally introduced by Bandt and Pompe.\citep{Bandt2002} 
The  entropy is a robust and computationally efficient metric for quantifying the
complexity and stochasticity of time series data and possesses several advantageous properties: it is invariant under monotonic
transformations, is robust to observational noise, and is 
easy to compute. In the context of dynamical systems, it has been
successfully employed to differentiate between periodic, chaotic,
and stochastic behaviors. When interpreted as a measure of stochasticity,
a higher value of the permutation entropy typically indicates increased randomness
and reduced temporal correlations. 

The foundational concept of permutation entropy is based on the
analysis of $\emph{ordinal patterns}$, which are defined by the relative
rankings of values within delay-embedded vectors. From the
iterative process, we can construct a real-valued time series representing
the differences in horizontal velocity. 
\begin{equation}
\left\{ \dv_{x_{i}}\right\} _{i=1}^{N_{\rm col}}=\left\{ \dv_{x_{1}},\dv_{x_{2}},\ldots,\dv_{x_{N_{\rm col}}}\right\} .
\end{equation}
From this series, delay-embedded vectors of dimension $d$ are formed
as
\begin{align}
\delta\vec{v}_{1}^{(d)}&=\left(\dv_{x_{1}},\dv_{x_{2}},\hdots,\dv_{x_{d}}\right) \nonumber \\
\delta\vec{v}_{2}^{(d)}&=\left(\dv_{x_{2}},\dv_{x_{3}},\hdots,\dv_{x_{d+1}}\right) \nonumber \\
 &\vdots   \nonumber \\
\delta\vec{v}_{N_{\rm col}-d+1}^{(d)}&=\left(\dv_{x_{N_{\rm col}-d+1}},\dv_{x_{N_{\rm col}-d+2}},\hdots,\dv_{x_{N_{\rm col}}}\right),
\label{embedded}
\end{align}
These embedded vectors serve as the basis for extracting ordinal patterns,
which are then used to compute the permutation entropy. 

Each vector $\delta\vec{v}_{\ell}^{(d)}$ is mapped to a permutation
$\pi_{\ell}$ that encodes the relative ordering of its components. For
example, if the entries of the vector satisfy $\Delta v_{x_{\ell}}<\dv_{x_{\ell+1}}<\cdots<\dv_{x_{\ell+(d-1)}}$,
the associated permutation is $\pi_{\ell}=(1,2,\ldots,d)$. Any other
ordering of the entries results in a different permutation. 
The permutation $\pi_{\ell}$ can be obtained from the $\texttt{argsort}$ function in Python, which returns the indices that  sort the vector in
ascending order
\begin{equation}
\pi_{\ell}=\texttt{argsort}\left(\delta\vec{v}_{\ell}^{(d)}\right).\label{argsort}
\end{equation}
From the embedded vector in Eq.~\eqref{embedded}, the function $\texttt{argsort}\left(\delta\vec{v}_{\ell}^{(d)}\right)$
creates a list of tuples: indexed\_values = $\left(\left(\dv_{x_{\ell}},1\right),\left(\dv_{x_{\ell+1}},2\right),\ldots,\left(\dv_{x_{\ell+(d-1)}},d\right)\right)$.
It then sorts indexed\_values by the first element of each tuple
(the value). Finally, it extracts and returns the list of second elements
(the original indices), generating the permutation $\pi_{\ell}$. 

The full set of ordinal patterns, denoted by $\Pi(S^{(d)})=\{\pi_{1},\pi_{2},\ldots,\pi_{N_{\rm col}-d+1}\}$,
forms the empirical basis for estimating the probability distribution
over $d!$ possible permutations, which in turn is used to compute
the normalized permutation entropy $H_{d}$. The relative frequency
of each permutation $\pi_{\ell}\in\Pi(S^{(d)})$ defines a probability
distribution over the $d!$ possible ordinal patterns. 
\begin{equation}
p_{\ell}=\frac{\text{Number of occurrences of }\pi_{\ell}\text{ in }\Pi(S^{(d)})}{N_{\rm col}-d+1}.\label{probality}
\end{equation}
Using this empirical distribution, the \emph{normalized permutation
entropy} $H_{d}$ is defined as
\begin{equation}
H_{d}=-\frac{1}{\log(d!)}\sum_{\ell=1}^{d!}p_{\ell}\log p_{\ell}.\label{PermEntrop}
\end{equation}
The value of $H_{d}$ lies within the interval $[0,1]$. Values close
to zero indicate regular or deterministic dynamics, whereas values approaching
one suggest high complexity or stochastic behavior. A typical classification
is summarized in Table~\ref{tab:range}.

\begin{table}[h]
\centering
\centering{}
\begin{tabular}{ll}
\hline 
\textbf{Range of $\boldsymbol{H_{d}}$} & \textbf{Interpretation}\tabularnewline
\hline 
$H<0.6$ & Regular or deterministic dynamics\tabularnewline
$0.6\leq H<0.9$ & Chaotic or noisy deterministic behavior\tabularnewline
$0.9\leq H\leq1.0$ & Likely stochastic or random process\tabularnewline
\hline 
\end{tabular}
\caption{\label{tab:range}Heuristic
thresholds of the permutation entropy commonly used to classify the degree of randomness in
a time series.}
\end{table}

As an  example, consider a time series of length
$N_{\rm col}=6$: $\{1,2,4,3,5,6\}$. Let us construct a sequence of embedded
vectors with dimension $d=3$: 
\begin{equation}
S^{(3)}=\left\{ \{1,2,4\},\{2,4,3\},\{4,3,5\},\{3,5,6\}\right\} .
\end{equation}
The $\texttt{argsort}$ operation results in 
{\scriptsize \begin{subequations}
\begin{eqnarray}
{ \{1,2,4\}} & { \rightarrow} & \negthickspace{ \{(1,\boldsymbol{1}),(2,\boldsymbol{2}),(4,\boldsymbol{3})\}\rightarrow\{(1,\boldsymbol{1}),(2,\boldsymbol{2}),(4,\boldsymbol{3})\}\rightarrow(\boldsymbol{1},\boldsymbol{2},\boldsymbol{3})}\nonumber\\
{ \{2,4,3\}}\negthickspace & { \rightarrow} & \negthickspace{ \{(2,\boldsymbol{1}),(4,\boldsymbol{2}),(3,\boldsymbol{3})\}\rightarrow\{(2,\boldsymbol{1}),(3,\boldsymbol{3}),(4,\boldsymbol{2})\}\rightarrow(\boldsymbol{1},\boldsymbol{3},\boldsymbol{2})}\nonumber\\
{ \{4,3,5\}}\negthickspace & { \rightarrow} & \negthickspace{ \{(4,\boldsymbol{1}),(3,\boldsymbol{2}),(5,\boldsymbol{3})\}\rightarrow\{(3,\boldsymbol{2}),(4,\boldsymbol{1}),(5,\boldsymbol{3})\}\rightarrow(\boldsymbol{2},\boldsymbol{1},\boldsymbol{3})}\nonumber\\
{ \{3,5,6\}}\negthickspace & { \rightarrow} & \negthickspace{ \{(3,\boldsymbol{1}),(5,\boldsymbol{2}),(6,\boldsymbol{3})\}\rightarrow\{(3,\boldsymbol{1}),(5,\boldsymbol{2}),(6,\boldsymbol{3})\}\rightarrow(\boldsymbol{1},\boldsymbol{2},\boldsymbol{3})}
\nonumber \end{eqnarray}
\end{subequations}}
In the pairs in parentheses $(a,{\bf b})$, $a$ represents the number in the list and ${\bf b}$ the order in the list. Initially, in the unordered list the order is the same for all vectors. Then the order is changed based on the values of $a$ in each vector so that the final result shows the ordering in each vector. 
Therefore the set of observed permutations is
\begin{equation}
\Pi(S^{(3)})=\left\{ (1,2,3),(1,3,2),(2,1,3),(1,2,3)\right\} .
\end{equation}
The complete set of permutations for $d=3$ consists of
\begin{equation}
\pi_{\ell}=\left\{ (1,2,3),(1,3,2),(2,1,3),(2,3,1),(3,1,2),(3,2,1)\right\} .
\end{equation}
We use Eq.~\eqref{probality} to find the  probabilities
\begin{subequations}
\begin{align}
p_{(1,2,3)} & =  \frac{2}{4},\quad p_{(1,3,2)}=\frac{1}{4},\quad p_{(2,1,3)}=\frac{1}{4},\\
p_{(2,3,1)} & =  0,\quad p_{(3,1,2)}=0,\quad p_{(3,2,1)}=0.
\end{align}
\end{subequations}
We substitute these values into the definition of permutation entropy, Eq.~\eqref{PermEntrop},
and obtain
\begin{equation}
H_{3}=-\frac{1}{\log(3!)}\left(\frac{1}{2}\log\frac{1}{2}+\frac{1}{4}\log\frac{1}{4}+\frac{1}{4}\log\frac{1}{4}\right)\approx0.580.
\end{equation}
Because $H_{3}<0.6$, this result suggests that the sequence is not
random and exhibits regular or deterministic behavior. 

By using this procedure, we can quantitatively assess the degree of stochasticity in the horizontal
component of the collision force and provide
a complementary perspective to the probabilistic analysis we have described
earlier. The results for three values of the embedding dimension $d$
and 6000 collisions are showed in Table~\ref{PermutTable}.

\begin{table}[h]
\begin{centering}
\begin{tabular}{cccc}
\hline 
$x_{0}$ & $d=3$ & $d=4$ & $d=5$\tabularnewline
\hline 
$-0.025$ & 0.997382 & 0.988037 & 0.97761\tabularnewline
$-0.012$ & 0.99849 & 0.99011 & 0.979425\tabularnewline
0.01 & 0.997483 & 0.986471 & 0.975847\tabularnewline
0.02 & 0.997045 & 0.985554 & 0.973457\tabularnewline
\hline 
\end{tabular}
\par\end{centering}
\caption{The permutation entropy computed for the bouncing ball  using four
different initial conditions and three values of the embedding dimension
$d$. Each time series was evolved for 6000 collisions. The entropy
values provide a quantitative measure of the randomness in the sequence
of horizontal bounces.}
\label{PermutTable}
\end{table}

Although the underlying dynamics is inherently deterministic, the
progression of the iterative process gives rise to behavior that effectively
simulates a stochastic system, as evidenced by the permutation entropy
values reported in Table~\ref{PermutTable}.

\section{Diffusion Process}

Statistical moments are essential tools in characterizing the behavior
of stochastic processes, particularly in the study of transport and
diffusion. To gain deeper insight into this behavior, we extend our analysis
by increasing the number of initial conditions while maintaining
the same total energy. Consider an ensemble
of $M$  distinct initial positions uniformly distributed on the $x$-axis with identical energy per unit mass $E/m$. The iterative procedure generates
$M$ distinct sequences, with each sequence consisting of $N_{\rm col}$ horizontal
positions of a particle, along with the corresponding velocity components
recorded at each collision with the surface, Eq.~\eqref{MAP3}. We focus
solely on the positional data so that the outcomes of the iterative processes
can be organized into a matrix of dimensions $M\times N_{\rm col}$, where each
row represents a distinct initial condition and each column corresponds
to a specific number of collisions,
\begin{equation}
X_{M\times N_{\rm col}}=\left(\begin{array}{cccc}
x_{1}^{(1)} & x_{2}^{(1)} & \cdots & x_{N_{\rm col}}^{(1)}\\
x_{1}^{(2)} & x_{2}^{(2)} & \cdots & x_{N_{\rm col}}^{(2)}\\
\vdots & \vdots & \ddots & \vdots\\
x_{1}^{(M)} & x_{2}^{(M)} & \cdots & x_{N_{\rm col}}^{(M)}
\end{array}\right).\label{ensemble1}
\end{equation}

To quantitatively characterize the diffusion of particles in the $x$-direction,
we analyze the
statistical distribution of particle positions over time. Given a
large ensemble of trajectories, the elements $x_{i}^{(j)}$  of $X_{M\times N_{\rm col}}$
 in Eq.~\eqref{ensemble1} are  used to construct a histogram.
This histogram provides a discrete approximation of the probability
density function, where the horizontal axis corresponds to the spatial
position and the vertical axis to the normalized frequency of occurrences.
The resolution of the histogram is determined by the choice of bin
width, which is chosen to balance the statistical noise and resolution. 

For each fixed $i$, the minimum and maximum values across the ensemble
are computed as
\begin{align}
x_{i}^{(\min)} & =  \min\{x_{i}^{(1)},\ldots,x_{i}^{(M)}\},\nonumber\\
x_{i}^{(\max)} & =  \max\{x_{i}^{(1)},\ldots,x_{i}^{(M)}\}
\end{align}
The data range at collision $i$ is then defined by
\begin{equation}
R_{i}=x_{i}^{(\max)}-x_{i}^{(\min)}.
\end{equation}
The number of histogram bins, $B$, plays a crucial role in determining
the resolution and interpretability of the resulting probability density
function. Various rules have been proposed for selecting $B$, including those by 
Sturges,\cite{sturges1926choice} Scott,\cite{scott1979optimal} and Freedman and Diaconis.\cite{freedman1981histogram}
In this work, we adopt  Sturger's rule, which defines the number
of bins as   
\begin{equation}
B=\left\lceil \log_{2}M\:+1\right\rceil ,
\end{equation}
where $\lceil \rceil$ denotes the ceiling function. 

Once $B$ is determined, the bin width, which is not a fixed number but depends on $i$,  is given by
\begin{equation}
w_i=\frac{R_{i}}{B}.
\end{equation}
The bin intervals at collision $i$ are then defined as
{\scriptsize\begin{eqnarray}
\mathcal{B}_{i}^{(\zeta)} & = & \left\{ \left[x_{i}^{(\min)}+(\zeta-1)w_i,\;x_{i}^{(\min)}+\zeta w_i \right)\;|\;\zeta=1,2,\ldots,B-1\right\} \nonumber \\
 &  & \cup\left\{ \left[x_{i}^{(\min)}+(B-1)w_i,\;x_{i}^{(\max)}\right]\right\} .
\end{eqnarray}}
For each bin $\zeta\in\{1,\ldots,B\}$, we count the number
of values in the set $\{x_{i}^{(1)},x_{i}^{(2)},\ldots,x_{i}^{(M)}\}$
that fall into the corresponding interval $\mathcal{B}_{i}^{(\zeta)}$.
This procedure yields the frequency count 
\begin{equation}
f_{i}^{(\zeta)}=\textrm{Count}\left(x_{i}^{(j)}\in\mathcal{B}_{i}^{(\zeta)}\;|\;j=1,\ldots,M\right).
\end{equation}
To convert these frequencies into a normalized probability density,
we define
\begin{equation}
p_{i}^{(\zeta)}=\frac{f_{i}^{(\zeta)}}{M  w_i}. \label{pn}
\end{equation}
Equation~\eqref{pn} ensures that the estimated probability density is
normalized, and thus $p_{i}^{(\zeta)}$ is the probability
 of finding a particle at the mean position $x_{i}^{(\zeta)}=x_{i}^{(\min)}+(\zeta-1/2)w_i$
at collision $i$. 

We consider an ensemble of initial horizontal positions $x_{0}$, uniformly distributed over the interval $[-3 \times 10^{-2},\, 3 \times 10^{-2}]\,\mathrm{m}$, with a spacing of $\Delta x_{0}=1.5 \times 10^{-5}\,\mathrm{m}$, resulting in a total of $M=4001$ distinct initial conditions. This ensemble may be interpreted as a collection of 4001 non-interacting particles, each initialized with the same specific energy $E/m=60.0\,\mathrm{J/kg}$, but differing in their initial spatial positions.

\begin{figure}[h]
\centering
\includegraphics[width=0.95\columnwidth]{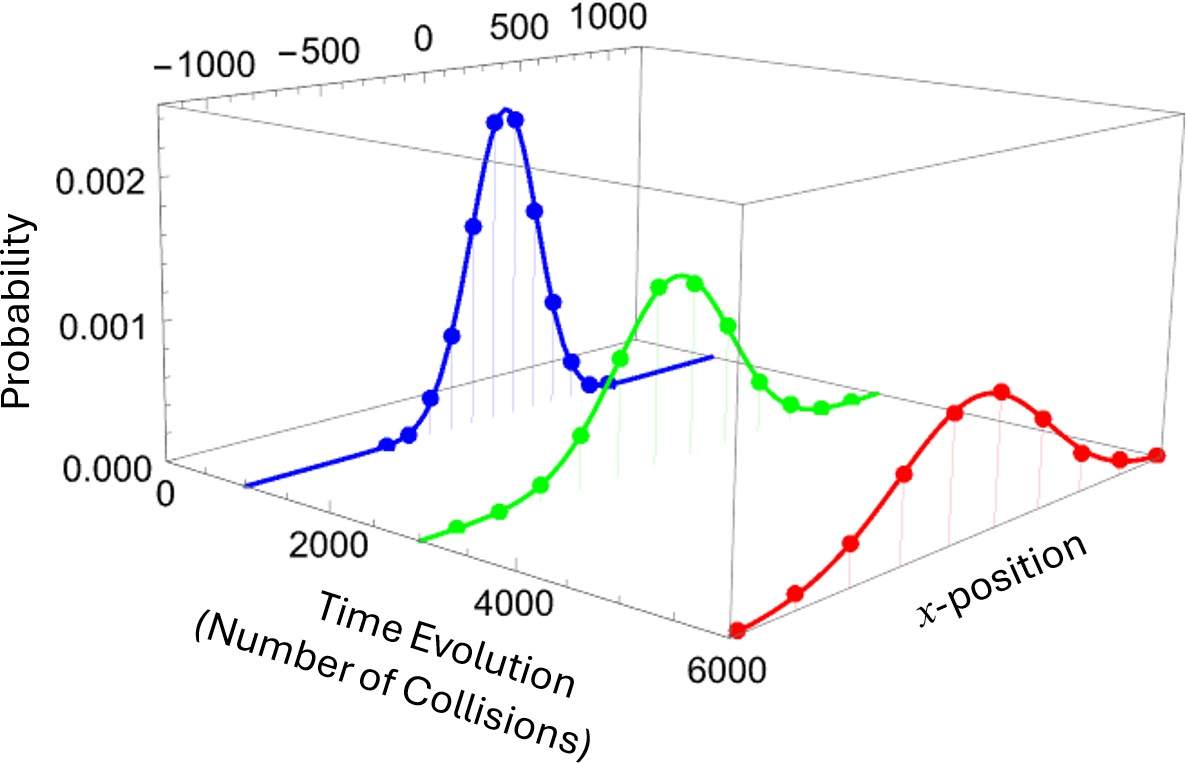}\caption{Three-dimensional visualization of the probability distributions of
horizontal positions at different numbers of collisions. The colored dots
represent the simulated data, and the continuous lines represent the
corresponding Gaussian fits. Blue corresponds to $i=1000$, green
to $i=3000$, and red to $i=6000$. The progressive spreading of the
distributions illustrates the diffusive behavior over time.}
\label{GaussianProbabil}
\end{figure}

We chose    $\alpha=100\,\mathrm{m}^{-1}$ and $\beta=0.02\,\mathrm{m}$ and computed 
 $N_{\rm col}=6000$ collisions   for each initial condition,
yielding the data matrix $X_{M\times N_{\rm col}}$. From this
matrix, the horizontal positions and their associated probabilities,
{\footnotesize \begin{equation}
\left\{ (x_{i}^{(\zeta)},\,p_{i}^{(\zeta)})\right\}_{\zeta=1}^{B}=
\left\{ (x_{i}^{(1)},\,p_{i}^{(1)}),\,(x_{i}^{(2)},\,p_{i}^{(2)}),\,\ldots,\,(x_{i}^{(B)},\,p_{i}^{(B)})\right\}  ,\label{coordGaussDistr}
\end{equation}}
are computed for $i=1000$, $i=3000$, and $i=6000$ collisions.
These results are graphically represented in Fig.~\ref{GaussianProbabil}.

The Gaussian behavior shown in Fig.~\ref{GaussianProbabil} is expected due to the stochastic-like
dynamics that emerge in the horizontal motion originating
from irregular momentum transfers during collisions with the wavy
surface. As a result, the diffusion process exhibits characteristics
analogous to a random walk. This resemblance  suggests that, under appropriate conditions and over sufficiently
long time scales, the probability distribution of horizontal displacements
converges to a Gaussian. Consequently, the histogram may be interpreted
as a sampling of a Gaussian distribution.

The fitting is performed by a
numerical minimization of the least squares error. The numerical fit begins by considering the functional form of the Gaussian
distribution
\begin{equation}
\phi (x;\mu,\sigma)=\frac{1}{\sqrt{2\pi\sigma^{2}}}\exp\left(-\frac{(x-\mu)^{2}}{2\sigma^{2}}\right),\label{ProbaDistribFunc}
\end{equation}
where $\mu$ and $\sigma>0$ represent the mean and standard deviation,
respectively. Given a normalized distribution $\left\{ (x_{i}^{(\zeta)},\,p_{i}^{(\zeta)})\right\} _{\zeta=1}^{B}$,
obtained from the simulated ensemble defined in Eq.~\eqref{ensemble1},
our objective is to determine the parameters $\mu$ and $\sigma$
that minimize the total squared deviation between the simulated probabilities
$p_{i}^{(\zeta)}$ and the model $\ensuremath{\phi (x_{i}^{(\zeta)};\mu,\sigma)}.$
We consider the objective function
\begin{equation}
S(\mu_{i},\sigma_{i})=\sum_{\zeta=1}^{B}\left[p_{i}^{(\zeta)}-\phi (x_{i}^{(\zeta)};\mu_{i},\sigma_{i})\right]^{2}.
\end{equation}

Because $S$ is nonlinear in both parameters, a gradient descent algorithm
is employed to numerically determine the optimal values. We let $\phi_{i}^{(\zeta)}=\phi (x_{i}^{(\zeta)};\mu_{i},\sigma_{i})$. 
The partial derivatives of $S$ with respect to $\mu_{i}$ and $\sigma_{i}$
are
\begin{align}
\frac{\partial S}{\partial\mu_{i}} & =-2\sum_{\zeta=1}^{B}\left(p_{i}^{(\zeta)}-\phi_{i}^{(\zeta)}\right) \phi_{i}^{(\zeta)}\frac{x_{i}^{(\zeta)}-\mu_{i}}{\sigma_{i}^{2}},\\
\frac{\partial S}{\partial\sigma_{i}} & =-2\sum_{\zeta=1}^{B}\left(p_{i}^{(\zeta)}-\phi_{i}^{(\zeta)}\right) \phi_{i}^{(\zeta)}\frac{\left(x_{i}^{(\zeta)}-\mu_{i}\right)^{2}-\sigma_{i}^{2}}{\sigma_{i}^{3}},
\end{align}
The parameters are then updated iteratively:
\begin{equation}
\mu\gets\mu-\eta \frac{\partial S}{\partial\mu},\quad\sigma\gets\max\left(\sigma-\eta \frac{\partial S}{\partial\sigma},\,\delta\right),
\end{equation}
where $\eta$ is a convergence factor, and $\delta>0$ ensures numerical
stability by preventing division by zero or negative variance. Because
the probabilities $p_{i}^{(\zeta)}$ are typically  the order of $10^{-3}$,
the partial derivatives are approximately $10^{-9}$. To facilitate
effective convergence, the value $\eta=10^{8}$ is employed.
The iteration continues until the convergence criteria are satisfied,
$|\Delta\mu|<\varepsilon\quad\text{and}\quad|\Delta\sigma|<\varepsilon.$ Here, $\varepsilon$ was chosen to be $10^{-3}$. The results for the mean and standard deviation are summarized in Table~\ref{TableMeanStandDev}. We observe that the mean remains close to zero and the standard deviation exhibits steady growth, consistent with a diffusive process.

\begin{table}[h]
\centering
\begin{tabular}{ccccccc}
\hline 
$i$ & 1000 & 2000 & 3000 & 4000 & 5000 & 6000\tabularnewline
\hline 
$\mu_{i}$ & 2.72 & 3.48 & -1.22 & 0.84 & 0.798 & 6.70\tabularnewline
$\sigma_{i}$ & 173 & 252 & 300 & 351 & 389 & 425\tabularnewline
\hline 
\end{tabular}
\caption{The mean position $\mu_{i}$ and standard deviation $\sigma_{i}$ of the
ensemble as functions of the number of collisions $i$. The values illustrate
the temporal evolution of the horizontal position distribution, where
the mean remains close to zero and the standard deviation grows with
time, indicating  diffusive behavior.}
\label{TableMeanStandDev}
\end{table}

In the context of an evolving stochastic processes, such
as random walks and particle diffusion, the analysis of statistical
moments plays a central role. Using the distribution in Eq.~(\ref{ProbaDistribFunc}), these moments provide quantitative measures
of the distribution's shape and spread, and can be computed by the
 expression
\begin{equation}
\langle x_{i}^{\nu}\rangle  
=  \frac{1}{\sqrt{2\pi\sigma_i^{2}}}  \intop_{-\infty}^{\infty}x^{\nu}\exp\left(-\frac{(x-\mu_i)^{2}}{2\sigma_i^{2}}\right)\,dx.
\label{moment}
\end{equation}

\begin{figure}[h]
\centering
\includegraphics[width=0.98\columnwidth]{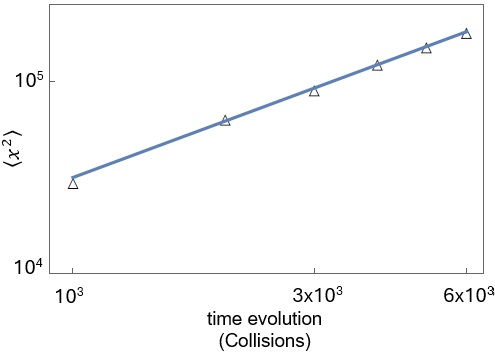}\caption{Time evolution of the second moment $\langle x_{i}^{2}\rangle$ as
a function of the number of collisions $i$. The data points (triangles)
correspond to the simulated values obtained from the ensemble, while
the solid line represents a power-law fit $\langle x_{i}^{2}\rangle=37.2\,i^{0.98}\approx 37.2\,i$. 
The value of the exponent close to unity indicates normal diffusion. }
\label{variance}
\end{figure}

The classification of diffusion regimes can be carried out by examining the scaling behavior of the second moment. We substitute $\nu = 2$ into the general moment expression, Eq.~(\ref{moment}), and obtain
\begin{equation}
\langle x_{i}^{2}\rangle = \mu_{i}^{2} + \sigma_{i}^{2} \approx \sigma_{i}^{2},
\end{equation}
because typically $\mu_{i}^{2} \ll \sigma_{i}^{2}$. Hence, the second moment $\langle x_{i}^{2}\rangle$ is effectively determined by the variance of the distribution. We use the
numerical results summarized in Table~\ref{TableMeanStandDev} to
 obtain the data points represented by triangles 
in Fig.~\ref{variance}. The solid line represents
a fit of the form
\begin{equation}
\langle x_{i}^{2}\rangle=37.2\,i^{\, 0.98} \approx 37.2\,i ,\label{diffusion}
\end{equation}
indicating that the second moment grows linearly with the $i$th collision.

The number of collisions (or iterations) $i$ plays the role of time and, in normal
diffusion, the mean squared displacement  grows linearly with
time as
\begin{equation}
\langle x_{i}^{2}\rangle\sim i^{\eta},
\end{equation}
with $\eta=1$. Deviations from this linear behavior signal the presence
of anomalous diffusion. For $0<\eta<1$,  the system is subdiffusive and
for $\eta>1$ the system is superdiffusive. The result obtained
in Eq.~\eqref{diffusion} shows that the system  with the parameters
$\alpha=100\,{\rm m}^{-1}$ and $\beta=0.02$\,m is consistent with  normal diffusive
behavior.

\begin{figure}[h]
\centering
\includegraphics[width=0.95\columnwidth]{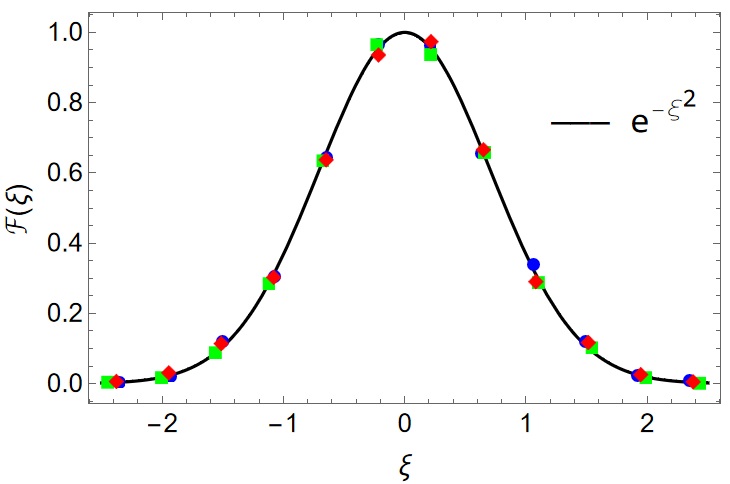}\caption{Scaled probability distributions for different times. The rescaling
is performed according to the transformation $\xi=x/\sqrt{2}\sigma_{i}$
and $\mathcal{F}(\xi)=\sqrt{2\pi}\sigma_{i}\phi (x;0,\sigma_i)$, ensuring that
all points collapse onto the same universal Gaussian curve $\exp(-\xi^{2})$.
This result confirms the approximate  diffusive behavior of the system
and validates the self-similarity of the distribution over time.}
\label{GaussianFinal}
\end{figure}

Finally, if the mean is set to zero, the distribution in Eq.~(\ref{ProbaDistribFunc})
can be rescaled by defining the Gaussian function
\begin{equation}
\mathcal{F}(\xi)=\exp(-\xi^{2}),
\end{equation}
which is related to the function $\phi (x;0,\sigma)$  by the change of variable $\xi=x/\sqrt{2}\sigma$ so that
\begin{equation}
\mathcal{F}(\xi)=\mathcal{F}\left(\frac{x}{\sqrt{2}\sigma}\right)=\sqrt{2\pi}\sigma \phi (x;0,\sigma).
\end{equation}
Therefore, if we scale each position and probability in the list (\ref{coordGaussDistr})
as
\begin{equation}
\left\{ \left(\frac{x_{i}^{(1)}}{\sqrt{2}\sigma_{i}},\,\sqrt{2\pi}\sigma_{i}\,p_{i}^{(1)}\right),\,\ldots,\,\left(\frac{x_{i}^{(B)}}{\sqrt{2}\sigma_{i}},\,\sqrt{2\pi}\sigma_{i}\,p_{i}^{(B)}\right)\right\} ,
\end{equation}
then all the points shown in Fig.~\ref{GaussianProbabil}
 collapse onto the same Gaussian curve $\exp(-\xi^{2})$, as illustrated
in Fig.~\ref{GaussianFinal}.

\section{Conclusion and Further Numerical Experiments}

We have investigated the horizontal diffusion behavior of
a particle undergoing successive collisions with a sinusoidal  
surface. The analysis revealed
that the particle's dynamics along the $x$-axis resemble a stochastic
process akin to a random walk. This interpretation
is supported by the near equiprobability of rightward and leftward
bounces, and is further corroborated using the   permutation
entropy, which quantifies the level of randomness  in the trajectory
sequences.

By simulating an ensemble of initial conditions and evaluating the
evolution of the system over time, we constructed probability density
functions for the horizontal positions. The statistical analysis
of these probability
density functions demonstrated that the mean displacement remains centered
around zero, while the second moment (variance) grows approximately
as $\langle x^{2}(t)\rangle\sim t$, indicating diffusive behavior.

Several directions for future numerical investigations emerge from
the present results:

\begin{enumerate}
\item \textit{Effect of surface roughness}.\\
To investigate the role of surface roughness, vary the geometric parameter 
$\beta$, which sets the amplitude of the floor undulations. Smaller $\beta$
corresponds to smoother surfaces and may enhance long ballistic flights. 
For each chosen $\beta$, compute the mean-squared displacement
$\langle x^{2}(t)\rangle$,
averaged over an ensemble of trajectories as explained in the text. Use a log--plot and fit the 
scaling law $\langle x^{2}(t)\rangle\sim t^{\eta}$ and classify the transport as 
subdiffusive ($\eta<1$), normal diffusive ($\eta\approx 1$), or 
superdiffusive ($\eta>1$). Preliminary simulations for $\beta=0.002$ 
yield $\eta\simeq 1.05$, suggesting a slightly superdiffusive regime. 
Explore a range of $\beta$ values from $0.001$ to $0.1$, 
such as $[0.001,\,0.005,\,0.01,\,0.02,\,0.03,\dots,\,0.1]$, and search for a correlation between 
$\beta$ and $\eta$, Identify possible transitions between diffusion regimes, and discussing the physical mechanisms 
(e.g., persistence of horizontal motion) underlying the observed behavior.

\item \textit{Comparison of the solutions of the simplified $\Delta t_{i,i+1}$  and the full transcendental equation}.\\
Examine the difference between solving the simplified
collision condition, Eq.~\eqref{eq:timeSpend}, and the full transcendental equation, Eq.~\eqref{eq:Transc}, for the impact
dynamics. This investigation should assess both the numerical accuracy 
of the final results and the computational efficiency, particularly
for long-time simulations.

\item \textit{Inclusion of dissipative effects}.\\
A more realistic model incorporates inelastic collisions by introducing a
dissipation coefficient $\gamma\in(0,1]$, such that the normal component
of the reflected velocity becomes $\gamma v_{\perp}$ [see
Eq.~\eqref{map1}]. The case $\gamma=1$ corresponds to fully elastic
collisions as discussed in the text. Consider 
$\gamma=0.8$, and analyze the following:

\begin{enumerate}
  \item Compute the total kinetic energy $E(t)$ as a function of time,
  averaged over an ensemble of trajectories with the same initial
  conditions used in the elastic case. Present results on log--linear
  or log--log scales to clearly identify the dissipation dependence on $t$.
  \item Determine whether the mean-squared displacement
  $\langle x^{2}(t)\rangle$ continues to scale approximately as
  $t^{\eta}$ for long times. If so, estimate the effective exponent
  $\eta$ and compare it with the elastic case.
  \item Discuss whether the dissipative dynamics still supports a
  diffusive regime, or if the system tends toward localization/freezing
  at long times. Relate your findings to the physical role of $\gamma$.
\end{enumerate}
\end{enumerate}

These  problems can enhance the understanding of the interplay
between geometry, energy dissipation, and stochastic-like transport
in dynamical systems exhibiting deterministic chaos. Such understanding
could be relevant for modeling transport phenomena in granular media,
corrugated surfaces, and engineered nanostructures where diffusion
emerges from complex microscopic interactions. 

\begin{acknowledgments}

The author acknowledges the support of the Coordination for the Improvement of Higher Education Personnel (CAPES).
\end{acknowledgments}

\appendix*

\section{The Secant Method for Root-Finding\label{sec:The-Secant-Method}}

In cases where the derivative of a function is difficult to evaluate or may vanish, the secant method provides a robust alternative to the Newton–Raphson method. Comprehensive descriptions of both methods are available in Refs.~\onlinecite{sauer2018numerical,sastry2012introductory,press2007numerical}.  The secant method is an iterative, derivative-free approach for approximating a root of a nonlinear equation $f(x) = 0$, relying on successive linear interpolations between function values.

Given two initial approximations $x_{0}$ and $x_{1}$, the method
constructs a secant line through the points $(x_{0},f(x_{0}))$ and
$(x_{1},f(x_{1}))$. The root of this line, which serves as the next
approximation $x_{2}$, is given by
\begin{equation}
x_{i+1}=x_{i}-f(x_{i})\frac{x_{i}-x_{i-1}}{f(x_{i})-f(x_{i-1})}.\label{eq:secant}
\end{equation}
This process is repeated iteratively until convergence is achieved
according to a specified tolerance: 
\begin{equation}
|x_{i+1}-x_{i}|<\varepsilon.
\end{equation}

Compared to the Newton-Raphson method, which has quadratic convergence
but requires the evaluation of $f'(x)$, the secant method achieves
superlinear convergence of order  $\varphi\approx\frac{1+\sqrt{5}}{2}\approx1.618$.
Despite its slightly slower convergence, it is particularly useful
when the derivative is not readily available or is numerically unstable.

The secant method was implemented  in
regions where the derivative becomes small or vanishes, ensuring numerical
stability during the root-finding process. Convergence was typically
achieved within 10 to 30 iterations. 

\bibliographystyle{apsrev}
\addcontentsline{toc}{section}{\refname}\bibliography{references}

\end{document}